\documentclass[pra,aps,amsfonts,amsmath,superscriptaddress,
showpacs,floatfix]{revtex4}

\usepackage{amsfonts}
\usepackage{amsmath}
\usepackage{bm}
\usepackage{epsfig}
\usepackage{ulem}

\newcommand{\grho}{\mathcal{R}}

\newcommand{\rr}{\boldsymbol{r}}
\newcommand{\braket}[1]{\langle#1\rangle}
\newcommand{\ket}[1]{|#1\rangle}

\begin{document}

\title {\bf The isovector dipole strength in nuclei with extreme neutron excess}

\author{D. Pe\~na Arteaga} \affiliation{Institut de Physique Nucl\'eaire,
Universit\'e Paris-Sud, IN2P3-CNRS, F-91406 Orsay Cedex, France}

\author{E. Khan} \affiliation{Institut de Physique Nucl\'eaire, Universit\'e
Paris-Sud, IN2P3-CNRS, F-91406 Orsay Cedex, France}

\author{P. Ring} \affiliation{Physikdepartment, Technische Universit\"{a}t
M\"{u}nchen, D-85748, Garching, Germany}

\begin{abstract}
  The E1 strength is systematically analyzed in very neutron-rich Sn nuclei,
  beyond $^{132}$Sn until $^{166}$Sn, within the Relativistic Quasiparticle
  Random Phase Approximation. The great neutron excess favors the appearance of
  a deformed ground state for $^{142-162}$Sn.  The evolution of the low-lying
  strength in deformed nuclei is determined by the interplay of two factors,
  isospin asymmetry and deformation: while greater neutron excess increases the
  total low-lying strength, deformation hinders and spreads it. Very neutron
  rich deformed nuclei may not be as good candidates as stable spherical nuclei
  like $^{132}$Sn for the experimental study of low-lying E1 strength.
\end{abstract}


\maketitle

\section{Introduction}

Collective modes of excitation are an universal feature of
nuclei~\cite{har01}. Involving a large majority of nucleons, these
modes can provide crucial insight into exotic nuclear structure. For
instance, the pygmy mode is known to be a consequence of a vibration
involving the neutron skin~\cite{paa07} against a core composed of
neutrons and protons.  Originally the increase of the low-lying
strength was thought to be an exclusive phenomenon of heavy nuclei
with large isospin asymmetry. But very similar excitation patterns
have also been observed in light nuclei~\cite{AETAL1999,TETAL2001},
and in medium-heavy stable nuclei~\cite{wag08,sav06,sav08}. It has
been also thoroughly investigated within a variety of theoretical
tools~\cite{REIN1999,paa07,TU2001}, including the RHB+RQRPA in
spherical symmetry~\cite{PNVR2005,PVR2005,VNPR2004,PRNV2003}, with
different degrees of success in comparison with experimental data.

The next generation of RIB facilities, such as FAIR, Spiral2 or RIBF, will
provide a wealth of new data on nuclear structure under extreme conditions, for
example in nuclei closer to the drip line. Hence it is of interest to perform
theoretical investigations on nuclear excitations under such conditions as
extreme isospin or deformation. Specifically, the present work will focus on
the study the E1 response in very neutron-rich tin isotopes, from $^{132}$Sn
(where current experimental data stands) up to $^{166}$Sn.

Several experimental and theoretical studies available on the less
neutron-rich part of the isotopic chain already provide some
information on the character of the low-lying E1 response. In
general, it has been found that the low-lying E1 strength increases
with neutron excess~\cite{paa07}, and that the collectivity of the
excitations is rather high. Tin nuclei are spherical because of the
Z=50 shell closure. However it is believed that for strong neutron
excess (beyond A ${\sim}$ 140), the large diffuseness of the density
induces a weakening of the spin-orbit potential~\cite{dob94,LVPR.98},
and therefore the disappearance of the shell closure, allowing for
deformed nuclei. This means that in order to describe low energy
excitations in these very neutron-rich nuclei, beyond $^{132}$Sn, a
model including both pairing and deformation is required. More
generally, most nuclei in the nuclear chart are deformed, which
requires to investigate the interplay between the low-lying E1
strength and deformation when going towards the drip-line. The aim of
this work is to investigate the influence of deformation on the
low-lying E1 strength for neutron-rich nuclei.

The recently developed~\cite{pen08} relativistic deformed QRPA
approach (RQRPAZ), provides a viable microscopical framework where
deformation and pairing correlations are included in a fully
self-consistent fashion. It should also be noted that the evolution
of the giant dipole resonance with respect to neutron excess and
deformation has not been investigated yet in a self-consistent
microscopic framework. The manuscript is organized as follows: in the
first section of this work we first very briefly recall the RHB and
RQRPA formalism. In Section II we present the ground-state properties
of the $^{132-166}$Sn chain of nuclei as calculated with the
parameter set NL3~\cite{NL3}, and study the electric dipole response
in the GDR and PDR energy regions, concentrating on the low-lying E1
strength. The last section is devoted to a brief summary and the
conclusions.

\section{Deformed RHB and QRPA formalism}

A detailed discussion of deformed Relativistic Hartree Bogoliubov
(RHB) and Relativistic Random Phase Approximation for axially
deformed systems can be found in references~\cite{GRT.90}
and~\cite{pen08}, respectively. Therefore, only the basic theory will
be outlined, and details given where extensions, like the inclusion
of pairing at the RPA level, are specific to this work.

The starting point is an effective Lagrangian density
\begin{equation}
  \mathcal{L}=\mathcal{L}_{N}+\mathcal{L}_{m}+\mathcal{L}_{int}.
\label{e:lagdens}
\end{equation}
  $\mathcal{L}_{N}$ refers to the Lagrangian of the free nucleon
\begin{equation}
  \mathcal{L}_{N}=\bar{\psi}(i{\gamma}^{\mu}{\partial}_{\mu}-m){\psi},
\end{equation}
where $m$ is the bare nucleon mass and ${\psi}$ denotes the Dirac spinor.
$\mathcal{L}_{m}$ is the Lagrangian of the free meson fields and the
electromagnetic field
\begin{align}
  \mathcal{L}_{m} & =   \frac{1}{2}\partial_{\mu}\sigma\partial^{\mu}\sigma
                      - \frac{1}{2}m_{\sigma}^{2}{\sigma}^{2}
                      - \frac{1}{4}\Omega_{\mu\nu}{\Omega}^{\mu\nu}
                      + \frac{1}{2}m_{\omega}^{2}{\omega}_{\mu}{\omega}^{\mu}
                      \nonumber \\
                  &   - \frac{1}{4}\vec{R}_{\mu\nu}\vec{R}^{\mu\nu}
                      + \frac{1}{2}m_{\rho}^{2}\vec{\rho}_{\mu}\vec{\rho}^{\mu}
                      - \frac{1}{4}F_{\mu\nu}F^{\mu\nu}
\end{align}
with the corresponding masses $m_{\sigma}$, $m_{\omega}$, $m_{\rho}$.
The interaction Lagrangian $\mathcal{L}_{int}$
is given by minimal coupling terms
\begin{align}
\mathcal{L}_{int}= & -g_{\sigma}\bar{\psi}\sigma\psi
             -g_{\omega}\bar{\psi}\gamma^{\mu}\omega_{\mu}\psi \nonumber \\
           & -g_{\rho}  \bar{\psi}\gamma^{\mu}\vec{\tau}\vec{\rho}_{\mu}\psi
              -e\bar{\psi}\frac{1}{2}(1-\tau_{3})\gamma^{\mu}A_{\mu}\psi
\end{align}
where $g_{\sigma}$, $g_{\omega}$, $g_{\rho}$ and $e$ are the
respective coupling constants for the ${\sigma}$, ${\omega}$,
$\vec{\rho}$ and photon fields.  However, this simple linear
interaction does not provide a quantitative description of complex
nuclear systems, and an effective density dependence needs to be
introduced. Historically, the first~\cite{BB.77} was the inclusion of
non-linear self-interaction terms in the meson part of the Lagrangian
in the form of a quartic ${\sigma}$ potential
\begin{equation}
  \frac{1}{2}m_{\sigma}^{2}{\sigma}^{2}
  + \frac{g_{2}}{3}\sigma^{3}+\frac{g_{3}}{4}\sigma^{4}
\end{equation}
which includes the non-linear ${\sigma}$ self-interactions with two
additional parameters $g_{2}$ and $g_{3}$. This particular form of
the non-linear potential has become standard in applications of RMF
functionals, although additional non-linear interaction terms, both
in the isoscalar and isovector channels, have been considered over
the years~\cite{Bod.91,TM1,SFM.00,HP.01}.

Two other approaches, of more recent development, can also be found
in the literature, based on the introduction of the density
dependence directly in the coupling
constants~\cite{TW.99,DD-ME1,DD-ME2} and on the expansion of the
meson propagators into zero-range couplings and gradient correction
terms~\cite{BMM.02,NVLR.08a,NVLR.08b}. In particular, the description
of the isovector channel has been greatly improved, which is
important for the quantitative description of neutron skins and
low-lying excitations. For example, non-linear density functionals
are known to consistently overestimate neutron
skins~\cite{DD-ME1,SR.92}. However, this discussion will be
restricted only to non-linear density functionals with the NL3
parameter set:  the increased computational requirements needed, for
example, to do the same calculations with density functionals with
density-dependent coupling constants, are such that they make
unfeasible to compute as big isotopic chain as $^{132-166}$Sn, that
includes many deformed nuclei, at the time of this writing. However,
work in this direction is in progress.

The Hamiltonian density can be derived from the Lagrangian density of
Eq.~\eqref{e:lagdens} as the (0,0) component of the energy-momentum
tensor, leading the to the energy functional $E[\hat{\rho},\phi]$
(for details see Ref.~\cite{VALR.05}).
\begin{align}
  E_{\mathrm{RMF}}[\hat{\rho},\phi] & =
   \mathrm{Tr}[(-i\mathbf{\alpha}\mathbf{\nabla}+\beta m)\hat{\rho}]
  +\sum_m\mathrm{Tr}[(\beta\Gamma_{m}\phi_{m})\hat{\rho}]\nonumber \\
  & \pm\frac{1}{2}\sum_m{\int}d^{3}r
  \left[({\partial}_{\mu}{\phi}_{m})^{2}+m_{m}^{2}\right],
  \label{e:efunctional}
\end{align}

In nuclei with open shells this simple covariant energy functional
fails, in general, to properly describe the nuclear many-body system.
It is thus necessary to introduce an additional field, the pairing
potential.  In order to include pairing correlations in a microscopic
way, the meson fields need to be quantized, so as to gain one meson
exchange two-body forces~\cite{KR91} (i.e terms of the form
$\psi^{\dagger}\psi^{\dagger}$). It is possible, however, to follow a
phenomenological approach and introduce a generalized Valatin density
$\hat{\grho}$~\cite{Vala61}

\begin{equation}
  \hat{\grho} =
  \left( \begin{array}{cc}
     \hat{\rho} & \hat{\kappa} \\
    -\hat{\kappa}^{\star} & 1- \hat{\rho}^{\star}
  \end{array} \right)
\end{equation}
where $\hat{\rho}$ is the single-particle density and $\hat{\kappa}$ is the
pairing density, and extend the energy functional to additionally depend on
it
\begin{equation}
  E_{\mathrm{RHB}}[\hat{\rho},\hat{\kappa},\phi] =
  E_{\mathrm{RMF}}[\hat{\rho},\phi] + E_{\mathrm{pair}}[\hat{\kappa}]
\end{equation}
where, in general, the pairing energy density can be expressed as
\begin{equation}
  E_{pair}[\hat{\kappa}] = \frac{1}{4}
  \mathrm{Tr}[\hat{\kappa} V^{pp} \hat{\kappa} ]
\end{equation}
for some yet to be defined effective pairing interaction $V^{pp}$.

For computational efficiency reasons, the results presented in this
investigation have been obtained using a simple monopole-monopole
pairing interaction with a smooth cutoff window~\cite{Bon85}, which
can be written as
\begin{equation}
  V_{kl'k'l}^{pp} = - \frac{G}{2} \;
  \frac{\delta_{kl'}\delta_{k'l}}
  {
    \left[ (1+e^{(\varepsilon_{k } - w) / d})
        (1+e^{(\varepsilon_{k'} - w) / d})
    \right]^{\frac{1}{2}}
  }
\end{equation}
where $w$ is the pairing window, $d$ its diffuseness, and
$\varepsilon_k = \braket{ k | h^{\mathcal{D}} | k }$ are the
eigenvalues of the Dirac single-particle Hamiltonian. This leads to
the gap equation
\begin{equation}
     \Delta_k = - \frac{1}{2} \sum_{k'} V^{pp}_{k\bar{k}k'\bar{k}'}
     \frac{\Delta_{k'}}{E_{k'}}
\end{equation}
To study vibrational excitations, one introduces small harmonic
oscillations around the ground state generalized density and expands
the equation of motion ~\cite{Ring80} up to linear order, gaining the
RPA approximation, which in standard matrix form reads
\begin{equation}
  \left( \begin{array}{cc}
    A & B \\
    -B^{*} & -A^{*}
  \end{array} \right)
  \left( \begin{array}{c}
    X^{(\nu)} \\ Y^{(\nu)}
  \end{array} \right)
  =
  \Omega^{(\nu)}
  \left( \begin{array}{c}
    X^{(\nu)} \\ Y^{(\nu)}
  \end{array} \right)
  \label{e:rpaeq}
\end{equation}
where the $X^{(\nu)}$ refers to the forward and backward amplitudes
of the transition density.

The matrix elements of the residual interaction
$V_{kl^{\prime}k^{\prime}l}^{ph}$ are the second derivatives of the energy
functional with respect to the single particle density
\begin{equation}
  V_{kl^{\prime}k^{\prime}l}^{ph}=\langle kl^{\prime}|\hat{V}^{ph}|k^{\prime
  }l\rangle=\frac{\delta^{2}E_{\mathrm{RHB}}}
  {\delta\rho_{k^{\prime}k}\delta\rho_{ll^{\prime}}
  }\label{rpa:vphph}
\end{equation}
Similarly, the $pp$
matrix elements $V^{pp}_{kl^{\prime}k^{\prime}l}$ are given by the second
derivative of the energy functional, but this time with respect to the pairing
density $\hat{\kappa}$
\begin{equation}
\label{rpa:vpppp}%
V^{pp}_{kk'll'} = \braket{kk'| \hat{V}^{pp} |ll'} = \frac{ \delta^2
E_{\mathrm{RHB}}}
  {\delta \hat{\kappa}_{kk'} \delta \hat{\kappa}_{ll'} }
\end{equation}
The ph matrix elements are calculated using a Fourier-Bessel
decomposition in cylindrical coordinates~\cite{thes07,pen08}.  The
transition matrix elements from the ground state $\ket{0}$ to an
excited state $\ket{\nu}$ for a one-body external operator
$\hat{\mathcal{O}}$ are given by
\begin{equation}
  \braket{0 | \hat{\mathcal{O}} | \nu } = \sum_{kk'} \left( \mathcal{O}_{k'k}
  X^{(\nu)}_{kk'} + \mathcal{O}^{*}_{kk'}
  Y^{(\nu)}_{kk'} \right) \left( u_{k}v_{k'} + \tau v_{k}u_{k'} \right)
  \label{rpa:obo}
\end{equation}
where $\tau = \pm 1$ depending on the time reversal properties of the
operator $\hat{\mathcal{O}}$: $\tau = 1$ if it is positive under time
reversal, and $\tau = -1$ otherwise. The transition densities can be
calculated with the help of Eq. (\ref{rpa:obo}) using for
$\mathcal{O}$ the density operator in coordinate space
\begin{equation}
\hat{\rho}(\boldsymbol{r})=\sum_{i}\delta(\boldsymbol{r}%
-\boldsymbol{r}_{i})
\end{equation}
which can be written for an axially symmetric system
as~\cite{thes07,pen08}
\begin{equation}
\rho(r_{\perp},\varphi,z,t)=\rho_{0}(r_{\perp},z)+\left[  \delta\rho(r_{\perp
},z)e^{-i(K\varphi+\omega_{\nu}t)}+h.c\right]  \label{def:tddensityaxi}%
\end{equation}
The two dimensional quantities $\delta\rho(r_{\perp},z)$ will be
plotted when discussing intrinsic transition densities. Projection of
the transition densities to the laboratory frame is accomplished
using the \textit{needle approximation}~\cite{Ring80}, where the
radial part (for a given external operator angular momentum L) can be
calculated from the intrinsic transition densities as
\begin{equation}
\delta\rho_{L}(r)=\int d\cos\theta d\varphi\;\delta\rho(r_{\perp}%
,z)Y_{LK}^{\ast}(\theta,\varphi) \label{def:prjtrdeq}%
\end{equation}
The definition of the quadrupole deformation parameter $\beta$ used throughout
this work as a measure of deformation is
\begin{equation}
  \beta_{n,p,t} = \frac{4\pi}{3} \sqrt{ \frac{5}{16\pi} }
  N_{n,p,t}^{-1} R^{-2}_{0} \int d^{3}r \; r^{2} Y_{20} \rho_{n,p,t}(\rr)
  \label{defdef}
\end{equation}
where the sub-indices $n,p,t$ refer to neutron, proton and total densities
($\rho$), particle number ($N$) and deformation ($\beta$); and $R_{0} = 1.2
A^{1/3}$ (fm).

\section{The very neutron-rich Sn isotopes}

\subsection{Ground state and deformation}

Table~\ref{tab:01} contains some of the bulk ground-state properties
of the $^{132-166}$Sn isotopic chain, as calculated with covariant
DFT with the NL3 parameter set. Since there is only scarce mass data
for such neutron rich nuclei, the monopole pairing constants
$G_{n,p}$ of the energy density functional were, for a fixed pairing
window ($w=20$ MeV, $d=1$ fm), adjusted to reproduce the corresponding
pairing gaps calculated with the parameter set D1S in the Gogny
calculations of Ref.~\cite{Hilaire02}. For the diagonalization of the
Dirac Hamiltonian, an expansion in anisotropic harmonic oscillator
wave-functions with N=20 major shells was used . The calculated RHB
binding energies are in very good agreement with the two available
experimental values for $^{132,134}$Sn~\cite{aud03}: B/A = -8.355 MeV
and -8.276 MeV, respectively. For $^{132}$Sn it is not very
surprising since the NL3 parameter set was adjusted including
experimental data from this very same nucleus. However, the agreement
in the case of $^{134}$Sn provides a sound basis for our pairing
prescription. The results of table~\ref{tab:01} are also in very good
agreement with the calculations in Ref.~\cite{LRR.99}. Small
differences have their origin in the different treatment of pairing.

\begin{table}[tbp]
  \begin{center}
  \begin{tabular}{cccccccc}
    \toprule
            A   & E/A (MeV) & R$_{n}$ (fm) & ${\beta}_{n}$
      & R$_{p}$ (fm) & ${\beta}_{p}$ & R (fm) & ${\beta}$ \\
            \hline
             132 & -8.362 & 4.99 & 0.00 & 4.64 & 0.00 & 4.86 & 0.00 \\
             134 & -8.283 & 5.08 & 0.00 & 4.66 & 0.00 & 4.93 & 0.00 \\
             136 & -8.193 & 5.15 & 0.00 & 4.67 & 0.00 & 4.98 & 0.00 \\
             138 & -8.099 & 5.20 & 0.00 & 4.69 & 0.00 & 5.02 & 0.00 \\
             130 & -8.010 & 5.25 & 0.00 & 4.71 & 0.00 & 5.06 & 0.00 \\
             142 & -7.931 & 5.30 & 0.13 & 4.74 & 0.07 & 5.11 & 0.11 \\
             144 & -7.852 & 5.35 & 0.20 & 4.78 & 0.13 & 5.16 & 0.17 \\
             146 & -7.775 & 5.34 & 0.19 & 4.80 & 0.13 & 5.20 & 0.17 \\
             148 & -7.699 & 5.45 & 0.18 & 4.82 & 0.13 & 5.25 & 0.16 \\
             150 & -7.626 & 5.52 & 0.33 & 4.89 & 0.27 & 5.32 & 0.31 \\
             152 & -7.547 & 5.56 & 0.33 & 4.92 & 0.28 & 5.36 & 0.32 \\
             154 & -7.471 & 5.60 & 0.32 & 4.93 & 0.28 & 5.34 & 0.31 \\
             156 & -7.396 & 5.65 & 0.31 & 4.94 & 0.27 & 5.43 & 0.29 \\
             158 & -7.321 & 5.69 & 0.29 & 4.95 & 0.26 & 5.46 & 0.28 \\
             160 & -7.247 & 5.73 & 0.26 & 4.96 & 0.24 & 5.50 & 0.25 \\
             162 & -7.169 & 5.75 & 0.19 & 4.95 & 0.18 & 5.51 & 0.18 \\
             164 & -7.123 & 5.76 & 0.00 & 4.91 & 0.00 & 5.52 & 0.00 \\
             166 & -7.048 & 5.80 & 0.00 & 4.93 & 0.00 & 5.55 & 0.00 \\
            \hline
        \end{tabular}
    \end{center}
    \caption{Ground state bulk properties along the $^{132-166}$Sn
  isotopic chain.}
 \label{tab:01}
\end{table}

\begin{figure}[ptbh]
  \includegraphics[width=8.6cm]{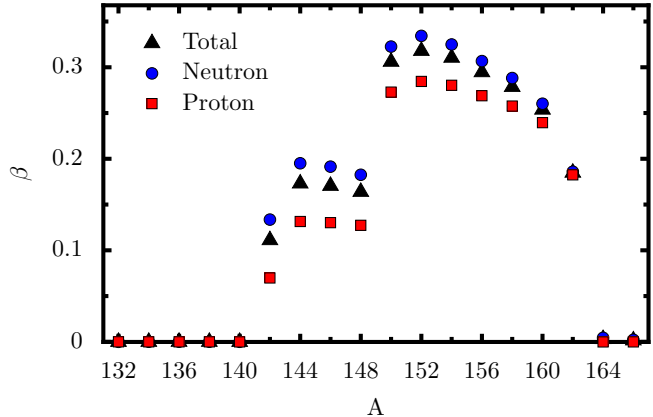}
  \caption{(Color online) Total, neutron and proton deformations.}
  \label{f:01}
\end{figure}

Fig.~\ref{f:01} shows the evolution of the quadrupole deformation
parameter $\beta$ (\ref{defdef}), for each nucleon species as well as
the total, with increasing neutron number. From $^{132}$Sn up to
$^{140}$Sn the mean field equations show a very well defined
spherical minimum in the potential energy surface. Coincidentally
with the fill up of the 1h$_{9/2}$ level, a minimum with axial
deformation appears for $^{142}$Sn. Deformation remains moderate
(${\beta}< 0.2$) until $^{150}$Sn, where the 2f$_{7/2}$ fills. From
then on, deformation varies smoothly until $^{164}$Sn, where tin
nuclei become spherical again due to the proximity of the N=126 shell
closure.

When deformation sets in at $^{142}$Sn, the neutron and proton
density profiles show different deformations for neutrons and
protons. This effect, that can also be found in calculations
performed with other relativistic models, and in non-relativistic
mean-field models using the Gogny force~\cite{webb3}, is
counter-intuitive if one considers the strong neutron-proton nuclear
interaction. However, one should also remember that even though the
proton-neutron interaction is strong enough to make stable semi-magic
nuclei spherical, neutron skins are still found for neutron rich
nuclei. It can thus be expected for deformed neutron-rich nuclei to
not exhibit the same quadrupole deformations for the proton and
neutron densities.  Coming back to Fig.~\ref{f:01}, from $^{142}$Sn,
as more neutrons are added, this difference in the $\beta$ parameter
is reduced to almost zero for the last deformed nucleus, $^{162}$Sn.
This reduction in the difference of deformations might be explained
by a progressive weakening of the spin-orbit interaction due to the
increased diffuseness of the neutron density as more weakly bound
neutrons are added.

These results compare rather well with predictions obtained using a
model with a more sophisticated pairing scheme. For example, a
deformed HFB approach with D1S Gogny functional shows that there is a
transition from spherical to deformed shape at $^{147}$Sn, and that
spherical symmetry is not restored until $^{163}$Sn~\cite{webb3}. It
is surprising, however, that the main differences in the predicted
deformation of the tin isotopes between both approaches happen at the
beginning of the chain, near the valley of $\beta$-stability, where
both forces, NL3 and DS1, were adjusted.

It should also be noted that the deformed tin nuclei are rather soft (i.e. the
potential energy surface is almost flat), both in the present study and in
results from~\cite{webb3}. The flatness of the potential energy surface can be
linked to quantum shape phase transitions~\cite{nik07}, which have a definite
effect on the vibrational response.  They may be studied specifically on the
isovector dipole response and deserve further consideration, even though it
falls outside the scope of the present manuscript.

\begin{figure}[ptbh]
  \includegraphics[width=8.6cm]{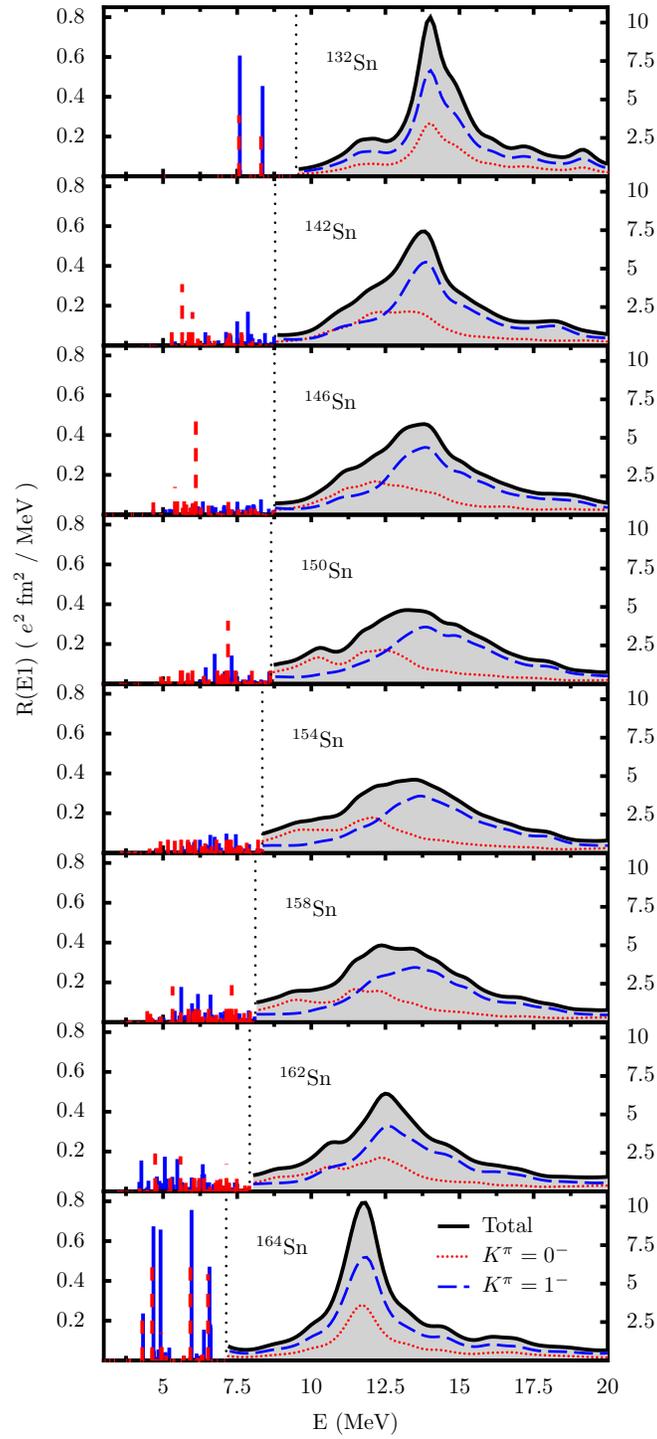}
  \caption{(Color online) B(E1) Isovector dipole response for selected Tin
  nuclei along the isotope chain $^{132-160}$Sn. The full response in the GDR
  region is indicated by the solid (black) line and shaded region.
  Contributions coming from the $K^{\pi}=1^{-}$ and $K^{\pi}=0^{-}$ modes are
  indicated with dashed (blue) and dotted (red) lines, respectively. The dotted
  vertical line marks the threshold that was considered as the upper bound for
  low-lying strength (see text for details). Below that threshold the response
  is not folded in order to be able to distinguish the details; solid (blue) and
  dashed (red) lines correspond to $K^{\pi}=0^{-}$ and $K^{\pi}=1^{-}$ peaks,
  respectively.}
  \label{f:02}
\end{figure}

\subsection{Response in the GDR region}

The RPA matrix equation was numerically solved by reduction to, and subsequent
diagonalization of, a non-Hermitian matrix of half the dimension (for technical
details please refer to Ref.~\cite{Papa07}).  The large configuration space
provided by a N=20 shells mean-field ground state makes unfeasible its complete
inclusion for the diagonalization of the RPA matrix. Thus, the number of
qp-pairs considered was truncated to approximately 16-20 thousand (depending on
the nucleus) out of $\approx$ 60 thousand , attending to energy and occupation
considerations. In particular, the energy cutoff for particle-hole and
antiparticle-hole pairs was set at 50~MeV and -1600~MeV, respectively. See
Refs.~\cite{DF.90,RMG.01} for a in-depth discussion about the need to include
anti-particle states in the calculation of the RPA response within the
\textit{no-sea} approximation. In addition, since the E1 response is calculated
within the same nucleus, the contribution of pairs that connect to neighboring
nuclei is small, and, in order to further reduce the configuration space to a
manageable dimension, the relation $u_{p}v_{h} > 0.05$ was also imposed on the
pairs.

This drastic truncation of the RPA configuration space induces a small lack of
self-consistency, and thus the translational spurious mode does not decouple at
exactly zero energy. Nevertheless, it was found reasonably close to zero for
all the nuclei in the isotopic chain, in the 0.5--0.8~MeV range. Moreover, it
was checked that the overlap of the spurious operator, in this case the total
linear momentum of the nucleus $\mathbf{P} = \sum_{i=1}^{N} \mathbf{p}_{i}$,
with the actual E1 response obtained in the calculations, was minimal: the
isoscalar components in the RPA transition densities do not come from
admixtures with the spurious mode, but correspond to the actual response of the
nuclear system to the E1 operator. This is very important in particular for the
study of the low-lying E1 strength, where one expects the appearance of mixed
isospin excitation modes like the pygmy resonance.

As mentioned before, there is no self-consistent RQRPA analysis of
the evolution of the GDR in deformed nuclei along an isotopic chain.
We shall therefore very briefly analyze this mode, although pairing
and deformation effects are expected to play a less important role
for these high energy collective modes.

Fig.~\ref{f:02} displays the evolution of the isovector dipole
response in Tin isotopes with increasing number of neutrons, from
$^{132}$Sn to $^{166}$Sn. The response is separated in two different
regions. The Giant Dipole Resonance region, to the right of the
dotted vertical line, shows the response obtained through
diagonalization of the RPA matrix folded with a 1 MeV Lorentzian. The
dotted (red) curve corresponds to excitations along the symmetry
axis, with $K^{\pi}=0^{-}$, while the dashed (blue) curve are those
perpendicular to the symmetry axis, with $K^{\pi}=1^{-}$. The total
added dipole response corresponds to the shaded region within the
solid (black) line. On the other hand, the low-energy region, to the
left of the dotted vertical line, shows the response without folding
in order to be able to distinguish details. Dashed (red) and solid
(blue) lines correspond to $K^{\pi}=0^{-}$ and $K^{\pi}=1^{-}$ peaks,
respectively.

It is not possible to choose a fixed energy threshold that separates the giant
dipole and pygmy regions along the whole isotopic chain. For very neutron rich
nuclei, the low-energy GDR tail comes down below any reasonable fixed threshold
that could be chosen for all the nuclei.  Therefore, a simple procedure was
adopted to separate the pygmy and giant dipole regions: first, a threshold
energy is chosen in a spherical nucleus where the separation is clear (in this
case it is in $^{132}$Sn at 9.5~MeV).  Then, for the rest of the nuclei in the
chain, the threshold is reduced the same amount the centroid of the E1 response
(as calculated using energy moments over the full energy range 0-30~MeV)
decreases, as compared to the chosen reference nucleus.  It was checked that
this simple procedure ensures that peaks with pygmy and giant dipole nature are
well separated in all cases, and that the relative values of the summed
low-lying strengths presented are stable to changes in the threshold energy.
This last point is particularly important since the following analysis and its
conclusions depend on the relative total low-lying strengths, and not in their
absolute value which, of course, is determined by the chosen energy threshold.

\begin{table}[tbp]
  \begin{center}
  \begin{tabular}{cccccc}
    \toprule
        A & $\bar{E}_{GDR}$ & ${\sum}_{GDR}$B(E1) & E$_{thres}$
    & ${\sum}_{PDR}$B(E1) \\
      & (MeV)                  & (\% TRK)            & (MeV)   & (\% TRK) \\
        \hline
        132 & 15.1 & 110.0 & 9.4 & 2.9 \\
        134 & 15.2 & 108.4 & 9.5 & 3.2 \\
        136 & 15.1 & 107.3 & 9.4 & 3.5 \\
        138 & 15.0 & 106.0 & 9.2 & 3.9 \\
        140 & 14.8 & 104.6 & 9.0 & 4.3 \\
        142 & 14.6 & 104.1 & 8.7 & 4.4 \\
        144 & 14.7 & 103.6 & 8.8 & 4.2 \\
        146 & 14.7 & 102.5 & 8.7 & 4.5 \\
        148 & 14.7 & 101.6 & 8.7 & 4.6 \\
        150 & 14.5 & 101.3 & 8.6 & 4.6 \\
        152 & 14.4 & 100.7 & 8.4 & 4.5 \\
        154 & 14.3 &  99.8 & 8.3 & 4.8 \\
        156 & 14.2 &  99.0 & 8.2 & 4.9 \\
        158 & 14.1 &  98.2 & 8.1 & 5.0 \\
        160 & 14.1 &  97.3 & 8.0 & 5.2 \\
        162 & 14.1 &  96.9 & 7.9 & 5.3 \\
        164 & 13.6 &  94.3 & 7.1 & 6.8 \\
        166 & 13.7 &  94.4 & 7.2 & 7.1 \\
    \hline
  \end{tabular}
  \end{center}
  \caption{Centroid energies of the GDR and PDR regions along the strength
  exhausted by each one in percents of the Thomas-Reihe sum rule (TRK).}
  \label{t:02}
\end{table}

Regarding the region above the pygmy threshold, as plotted in Fig.
~\ref{f:01}, two phenomena concerning the evolution of the GDR
strength are readily observable: i) The centroid position of the GDR
shifts to lower energies with increased particle number, and ii) the
width of the GDR increases with deformation and particle number.
These features are already known from many studies~\cite{har01}, but
it is checked here for the first time using a self-consistent
approach, and more generally on very neutron-rich nuclei. The net
effect is that the dipole response becomes very soft for deformed
nuclei.

\begin{figure}[ptbh]
  \includegraphics[width=8.6cm]{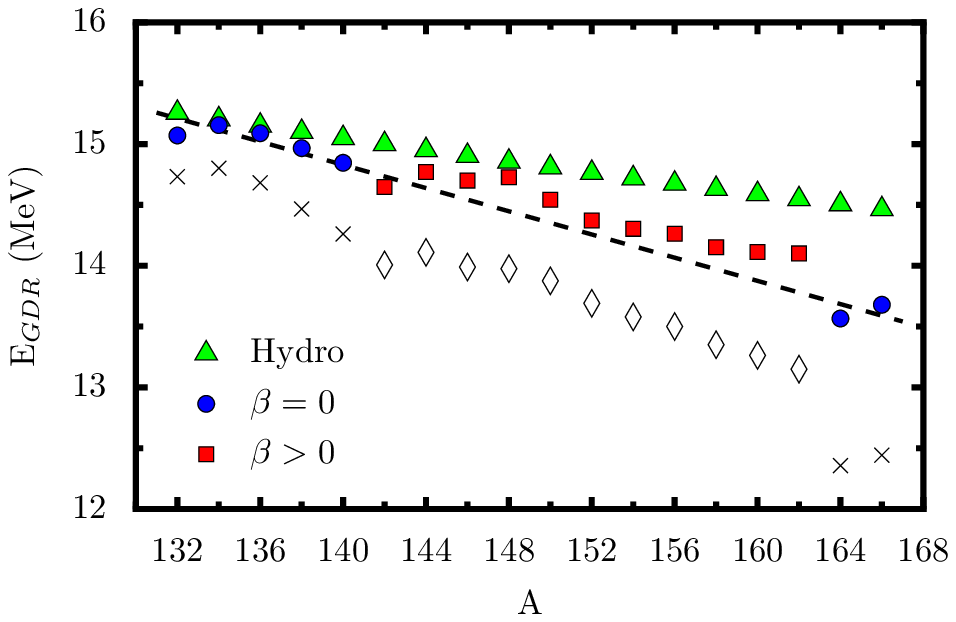}
  \includegraphics[width=8.6cm]{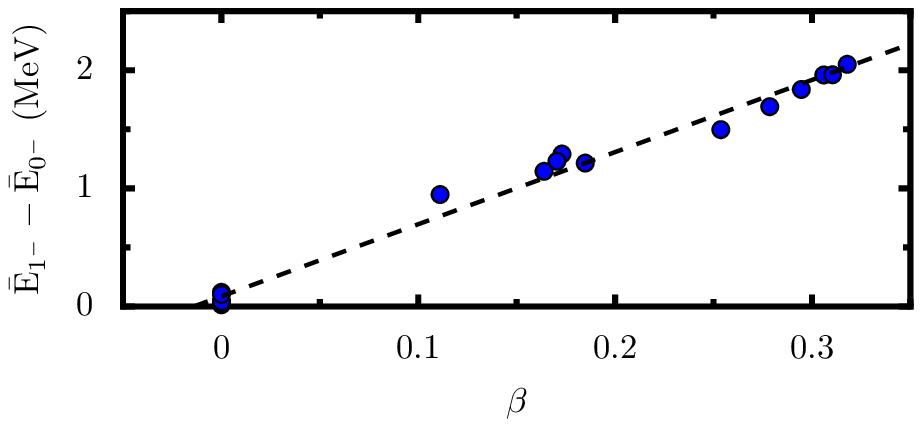}
  \caption{(Color online) Upper panel: GDR centroid position as predicted by
  hydrodynamical models, compared with this the results obtained in this work.
  The dashed line represents the least-squares fit to the calculated GDR
  position in spherical nuclei (blue circles) along the isotopic chain.
  Deformed nuclei are marked with (red) squares. The hollow crosses and
  diamonds refer to the centroid position of the spherical and deformed nuclei
  as calculated averaging over the whole energy range, from 0~MeV up to 30~MeV.
  Lower panel: GDR splitting versus the deformation parameter $\beta$.}
  \label{f:03} \end{figure}

In the upper panel of Fig.~\ref{f:03} the GDR centroid position dependence on
the mass number is plotted, as well as the predictions of the hydrodynamical
model~\cite{har01} (green triangles).  The RQRPAZ values (blue circles and red
squares) were calculated as the energy centroids of the response, from the
threshold energy (see Table~\ref{t:02}) up to 30 MeV. The dashed line indicates
a least-squares fit to the GDR position of spherical nuclei. As expected, the
differences increase when going away from the valley of stability. The crosses
and diamonds indicate the centroid QRPA energy as calculated when including the
whole energy interval, from 0~Mev to 30~MeV. Their deviation from the squares
and circles show that the low-lying strength plays an increasingly important
role with the addition of neutrons, i.e., the E1 distribution cannot be
considered to be composed of only a very well developed GDR peak.

Even though it falls outside the scope of this investigation, it would be
interesting to study the same isotopic chain using parameters sets and
functional forms of the covariant Lagrangian.  Measurements of the GDR could be
used to further constrain the isospin dependence of the density functional,
which would have direct applications to nuclear structure and astrophysical
problems~\cite{Khan.02,Khan.04}.

As expected, for spherical nuclei the excitation peaks for both the
$K^{\pi}=0^{-}$ and $K^{\pi}=1^{-}$ modes lie at the same energies
(Fig.~\ref{f:02}, $^{132}$Sn and $^{164}$Sn). However, once the nucleus becomes
deformed, they split. In principle, for prolate nuclei, as is the case for the
studied Tin isotopes, the strength coming from the $K^{\pi}=0^{-}$ mode should
appear at lower energies compared to the $K^{\pi}=1^{-}$ mode. An intuitive
argument to explain this phenomenon recalls a very simple picture where the
nuclear potential must be flatter (more extended) along the symmetry axis, and
thus it is more favorable energetically for the nucleons to oscillate in that
direction ($K=0$) than perpendicular to the symmetry axis ($K=1$), where the
nuclear potential is narrower~\cite{Dan.58}.

Since deformation is the cause of the GDR splitting, it is possible, in
principle, to relate the nuclear deformation with the energy separation of the
two $K$-modes. In the lower panel of Fig.  ~\ref{f:03} is plotted the GDR
splitting along the $^{132-166}$Sn isotopic chain versus the quadrupole
deformation parameter ${\beta}$.  There is an approximate linear relation
between the two parameters, as was already predicted by hydrodynamical
models~\cite{har01}.

To finish with this brief analysis of the dipole strength in the Giant Dipole
region, Table~\ref{t:02} also displays the strength exhausted by the GDR in
percents of the classical Thomas-Reihe-Kuhn (TRK) sum rule. The sharp decrease
in strength is constant along the isotope chain, proportional to the
enhancement of low-lying response, and no big effects due to deformation can be
observed. Obviously, for spherical nuclei the contribution to the response
strength coming from the $K^{\pi}=1^{-}$ mode is found to be double than that
of the $K^{\pi}=0^{-}$ mode. However, for deformed nuclei the share of strength
exhausted by the $K^{\pi}=0^{-}$ mode increases slightly up to 37\% for the
most deformed nucleus in the chain, $^{152}$Sn.

\begin{figure}[tbhp]
  \includegraphics[width=8.6cm]{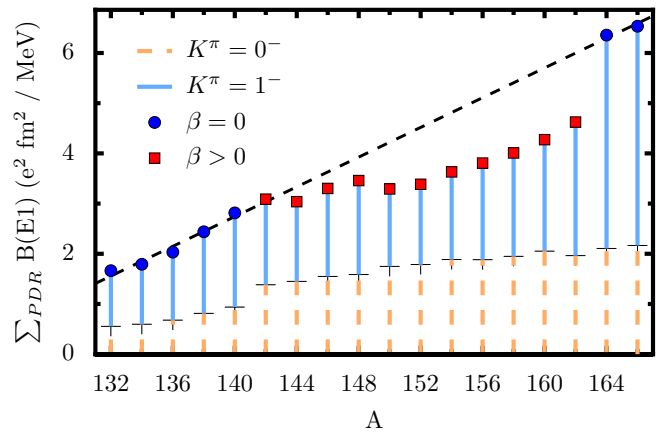}
  \caption{(Color online) Total pygmy strength versus the mass number. (blue)
  Circles and (red) squares denote spherical and deformed nuclei, respectively.
  The dashed (orange) vertical lines indicate the share of the total strength
  provided by the $K^{\pi}=0^{-}$ mode, while the solid (light blue) are the same
  but for the $K^{\pi}=1^{-}$ mode. }
  \label{f:04}
\end{figure}

\subsection{Low-lying dipole response}

The nature of the low-lying dipole response in neutron-rich nuclei is currently
under discussion~\cite{sav08,paa07}. All of the available microscopic models
predict an increase in dipole strength below 10~MeV for nuclei with high
isospin asymmetry, but differ on basic details, like excitation patterns or
collectivity of the low-lying modes. For example, for light nuclei, models
based on covariant DFT obtain the already well-known pygmy structure, while on
non-covariant models the response in this energy region is composed of
single-particle excitation peaks~\cite{paa07}. The low-lying E1 strength in
spherical Tin isotopes (from $^{114}$Sn to $^{140}$Sn) has been studied within
spherical RQRPA in Ref.~\cite{PNVR2005}, showing that with increasing neutron
number i) the share of strength exhausted by the low-lying part is generally
enhanced, ii) the centroid position of the low-lying modes decreases.

Nuclei with extreme isospin asymmetry such as $^{140}$Sn will be accessible in
next generation experimental facilities, and their study will hopefully shed
light on the nature of the low-lying dipole strength. It is thus relevant to
study the predictions of the different theoretical models on exotic nuclei. Of
course, as most of these nuclei are expected to be deformed, theoretical models
should make explicit allowance for the deformation degree of freedom. In this
regard, recent advances have been also made with non-relativistic models using
the Gogny~\cite{Sophie.08}, and Skyrme forces~\cite{Giai.08}.

\begin{figure}[tpbh]
  \includegraphics[width=8.6cm]{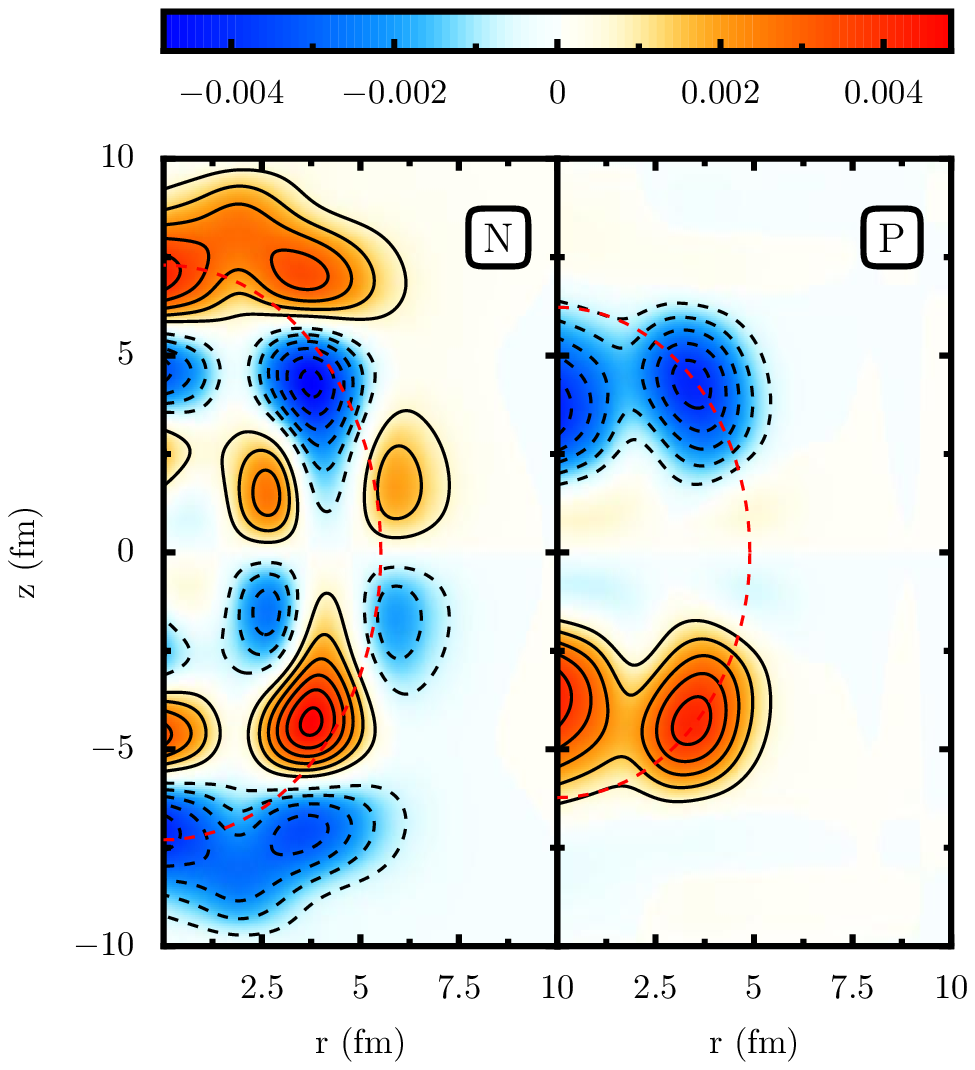}
  \includegraphics[width=8.6cm]{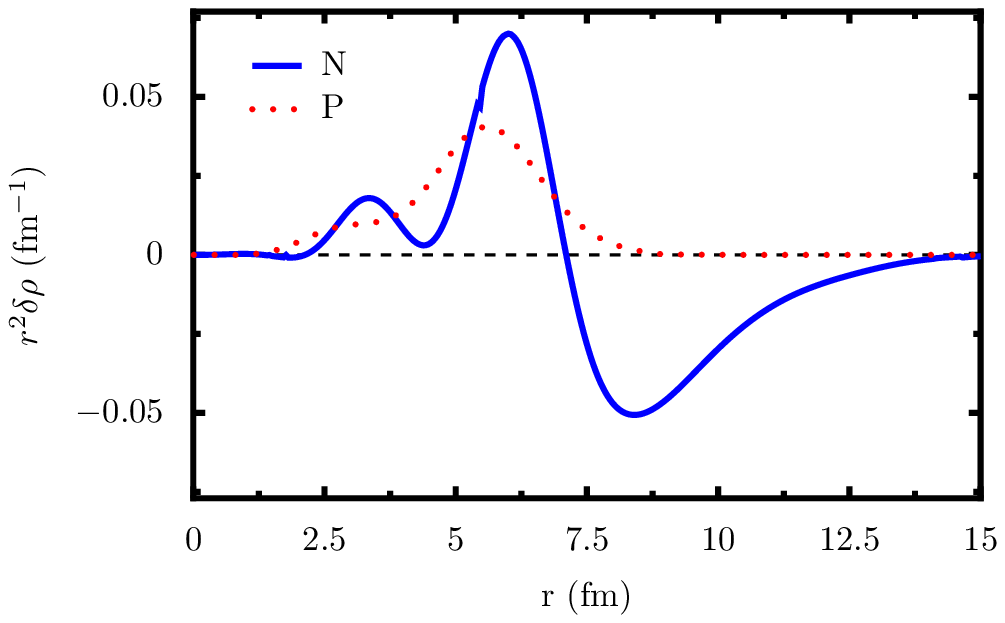}
  \caption{(Color online) Transition densities in the intrinsic frame
  and in the laboratory for the $0^{-}$- peak in
  $^{150}$Sn at 7.2~MeV. See text for details.}
  \label{fig:05}
\end{figure}

In the case of the present investigation within the RQRPA model in Tin
isotopes, the left panels in Fig.~\ref{f:02} show that the response in the
energy region below the threshold becomes more spread with increasing neutron
excess, up to the restoration of spherical symmetry in $^{164}$Sn, where it
shows again only a few contributing peaks. It is evident that for the most
deformed nuclei (i.e.  $\beta>0.2$), $^{150-162}$Sn, the low-lying strength
tends to be distributed into many different peaks, without any one of them
dominating the response. For mildly deformed nuclei ($\beta < 0.2$) there is
also significant Landau damping, even though a single (most cases) or several
(as for example in $^{162}$Sn) peaks clearly stand from the background
response. The overall conclusion is that deformation distributes the low-lying
strength, irregardless of the number of excess neutrons.

\begin{figure}[tpbh]
  \includegraphics[width=8.6cm]{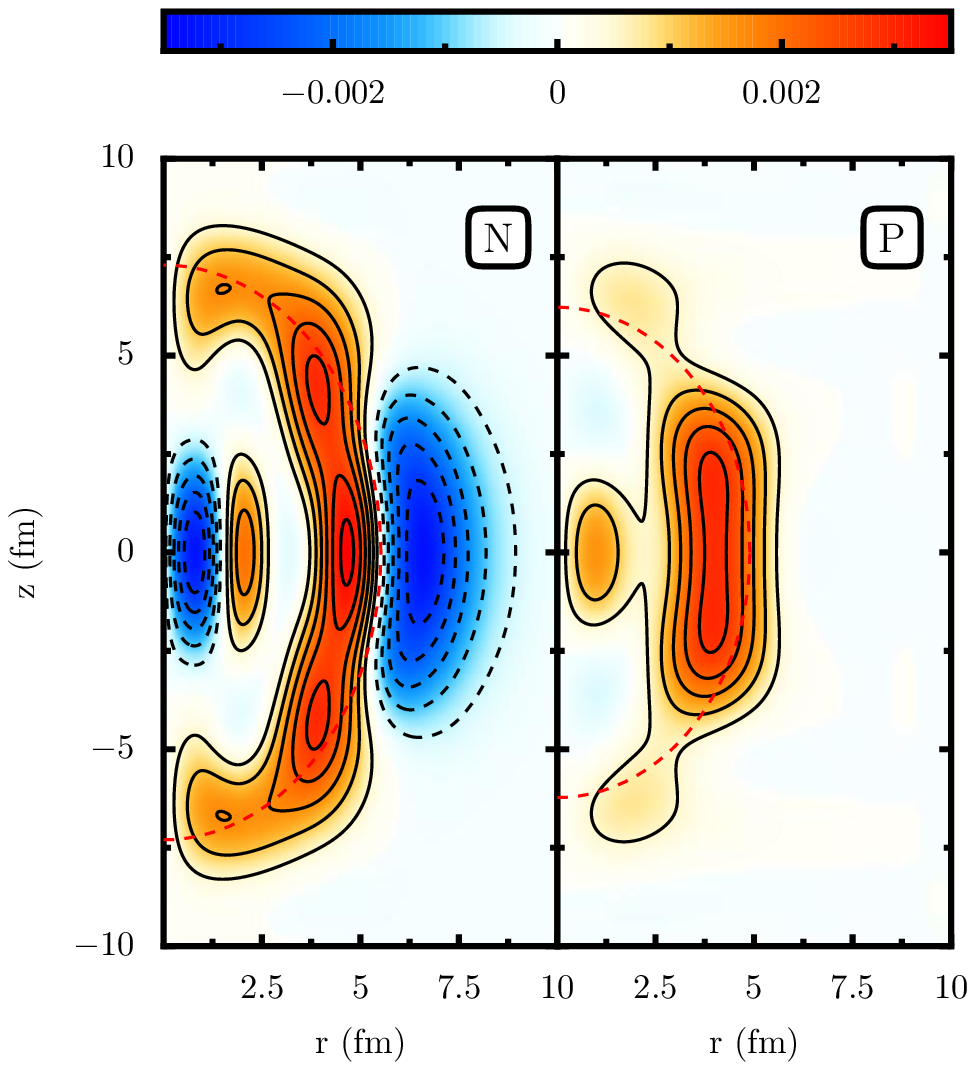}
  \includegraphics[width=8.6cm]{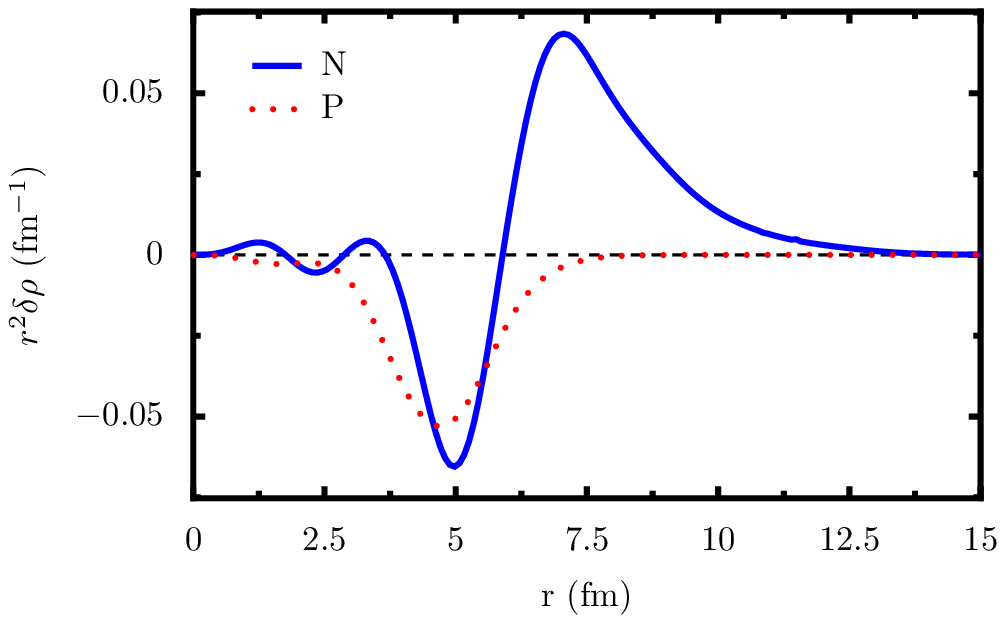}
  \caption{(Color online) Same as Fig.~\ref{fig:05} for the $K^{\pi}=1^{-}$ mode
  at 7.3~MeV in $^{150}$Sn.}
  \label{fig:06}
\end{figure}

To assert if, and to what extent, the low-lying response is in general affected
by deformation, it is interesting to concentrate on the total low lying
strength. Fig.~\ref{f:04} shows the summed response up to the threshold energy
versus the mass number for the whole isotopic chain. Again (blue) circles and
(red) squares refer to spherical and deformed nuclei, respectively. The dashed
line, which is a least squares fit to the data points for spherical nuclei,
clearly indicates that, all things being equal, for spherical nuclei the
low-lying strength increases almost linearly with neutron number.  It is
important to note, however, that studies within the RQRPA ~\cite{paa07} in the
spherical $^{100-132}$Sn nuclei, show that this trend is reversed  near a shell
closure, in this particular isotopic chain approaching the neutron number N=82.
Between $^{126}$Sn and $^{132}$Sn one finds a decrease of the PDR strength
(see~\cite{paa07} and references therein).

However, Fig.~\ref{f:04} also shows that the linear link between the addition
of neutrons and an increase in total low-lying strength is no longer kept for
deformed nuclei, where the growth is less pronounced.  Furthermore, for nuclei
where deformation most dramatically increases, from $^{148}$Sn to $^{150}$Sn,
and to the most deformed $^{152}$Sn, the summed low-lying strength even
decreases with the addition of two neutrons. As stated before, it has been
checked that this is not due to the particular energy threshold chosen, and
therefore, it has to be concluded that deformation quenches the dipole response
in the low-lying energy region.

\begin{figure}[ptbh]
  \includegraphics[width=8.6cm]{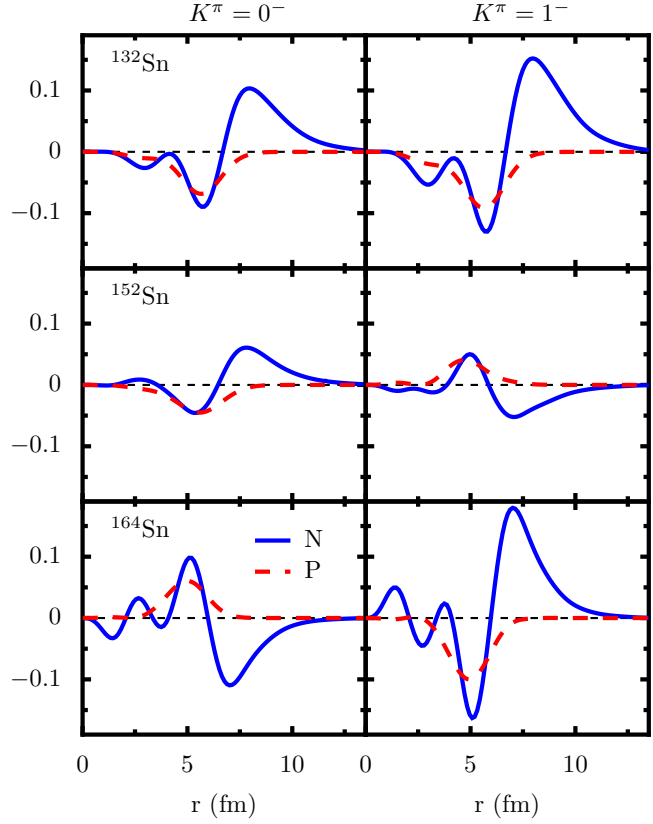}
  \caption{(Color online) Comparison of the pygmy mode's transition densities
  for the two K-modes between spherical and deformed nuclei. }
  \label{fig:07}
\end{figure}

The origin of this quenching can be further analyzed looking at the vertical
lines that mark each data point in Fig.~\ref{f:04}, which show the
decomposition of the strength into contributions coming from $K^{\pi}=0^{-}$
and $K^{\pi}=1^{-}$ modes.  For spherical nuclei the $K^{\pi}=1^{-}$ mode
carries two thirds of the total response, while the $K^{\pi}=0^{-}$ mode
provides the rest. However, for deformed nuclei in the isotopic chain the
contribution from the $K^{\pi}=0^{-}$ mode increases (in the case of the most
deformed nucleus $^{152}$Sn it reaches almost 60\% of the total), while that
from the $K^{\pi}=1^{-}$ mode significantly decreases. Thus, there is a
quenching of the $K^{\pi}=1^{-}$ mode and a smaller enhancement of the
$K^{\pi}=0^{-}$ mode, that leads to an overall quenching of the dipole strength
in the low-lying region.

Since this reduction comes mainly from the $K^{\pi}=1^{-}$ mode, it is likely
that there is a geometrical interpretation. It is important to realize,
however, that the validity of such a geometrical picture depends very much on
the collectivity and excitation structure of the RPA peaks.  In other studies
within spherical RQRPA~\cite{paa07} it has been found that the dominant
low-lying peaks in this region of the nuclear chart show a rather collective
structure, with a very characteristic pattern for the transition densities
which support the common interpretation of the low-lying strength as the pygmy
dipole resonance, a collective vibration of the skin of excess neutron against
a T=0 core.  Furthermore, a study using deformed RQRPA in $^{100}$Mo ($\beta
\approx 0.3$)~\cite{pen08b} shows that, at least for stable nuclei, deformation
does not destroy neither the excitation pattern nor the collectivity of the
PDR, but merely splits the response into different peaks for the different
$K^{\pi}$ modes.

To gain insight on the geometrical nature of the low-lying excitations one has
to look at the transition densities (\ref{def:tddensityaxi}), which are in the
case of axially deformed nuclei functions depending on the two coordinates
along ($z$) and perpendicular ($r$) to the symmetry axis. It has been found
that along the full chain of Tin isotopes under study, a pygmy-like structure
was present in the most dominant peaks in the low-lying energy region. As an
example, Figs.~\ref{fig:05} and~\ref{fig:06} show, in the upper panel, the 2D
transition densities of the $K^{\pi}=0^{-}$ (at 7.2~MeV) and $K^{\pi}=1^{-}$
(at 7.3~MeV) peaks in $^{152}$Sn. The upper panel of Fig.~\ref{fig:05} shows
the typical pygmy excitation pattern: inside the dotted line, the nuclear
interior, the transition densities for neutrons (left) and protons (right) are
in phase (same sign, i.e. same kind of contour lines and same color shading),
while in the surface region, outside the dotted (red) line, they are out of
phase in the case of neutrons, and non-existent for protons.  In the specific
case of Fig.~\ref{fig:05}, since it is the transition density for a
$K^{\pi}=0^{-}$ mode, the vibration takes place along a perpendicular of the
symmetry axis, i.e., the skin of neutrons (left panel, $-7fm<-z<-7fm$) is
concentrated at the caps of the prolate nuclear shape, along the symmetry axis.
On the other hand, on the upper panel of Fig.  ~\ref{fig:06} is plotted a
$K^{\pi}=1^{-}$ mode, and thus the excitation is along a perpendicular of the
symmetry axis, i.e. the neutron skin ($r>5$ fm) is concentrated around the
equator.

The 2D transition densities are referred to the intrinsic frame of reference,
where only the total angular momentum projection $K$ is a good quantum number,
i.e., they are expected to contain admixtures from all possible angular
momenta.  Since the E1 transition operator is a 1-rank tensor, it is expected
that the major contributions of the transition densities to the total response
come from the $I=1$ angular momentum. It is therefore interesting to obtain the
actual transition density that would be observed in the laboratory frame of
reference, after projection to $I=1$. The lower panels of Figs.~\ref{fig:05}
and~\ref{fig:06} show the radial part of the transition densities after such a
projection procedure. The pygmy pattern is easily recognizable, and is very
similar to those obtained in spherical systems: in the nuclear interior both
neutrons and protons oscillate in-phase, out-of-phase with the skin where only
neutrons contribute.

This pattern is observed for both $K$-modes for the non-negligible excitation
peaks in the low-lying region across the isotopic chain.  Fig.~\ref{fig:07}
shows, as an example, a comparison of the projected transition densities for
some selected spherical and deformed cases.  Obviously there is a direct
correlation between the share of strength coming from the different
$K^{\pi}$-modes and their transition density amplitudes. For example, for
$^{152}$Sn, where the contribution to the strength from the $K^{\pi}=0^{-}$
mode is slightly larger than that of the $K^{\pi}=1^{-}$ mode, the transition
densities show a similar magnitude.  This is in contrast to spherical nuclei
where it is trivial that the one third/two thirds ratio is exactly preserved
for both total response strength and transition density amplitude. In addition,
and even though the total response decreases, the share of strength from the
$K^{\pi}=0^{-}$ response increases with deformation (e.g. in Fig.~\ref{f:04}
from $^{140}$Sn to $^{142}$Sn).

The analysis of the single-particle structure of the peaks reveal that their
overall collectivity, measured as the number of contributing qp-pairs, is
comparable with those peaks in the GDR energy region.  In all the cases except
of $^{132}$Sn, the collectivity of the low-energy states is further enhanced
due to the opening of the neutron shell and the increased number of
two-quasiparticle configurations which contribute to the low-lying states. In
particular, the proton contribution does not exceed 10\% of the total in any of
the cases, and is usually around 3--4\%.  This is in good agreement with
previous RQRPA results in spherical systems, where similar collectivity and
proton contribution has been reported~\cite{paa07}.  Finally, expanding the
single-particle wave-functions in an anisotropic harmonic oscillator basis, one
finds that usually there is one dominant factor in the series, that can be used
qualitatively as a label. This provides insight into the kind of
single-particle excitations that compose each RPA peak (for details,
see~\cite{pen08}). In this regard, the pygmy resonance shares the same
structure as the GDR, in so far as both of them involve a change in major
quantum number of $\Delta N = 1$.

\begin{figure}[ptbh]
  \includegraphics{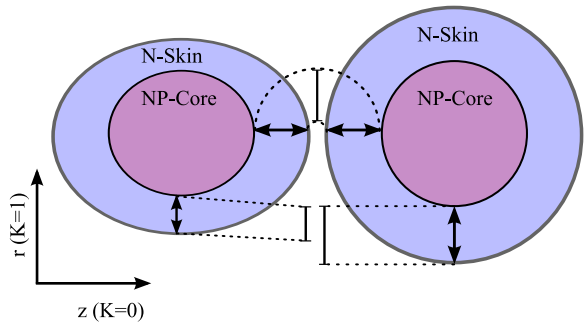}
  \caption{(Color online) Schematic diagram that shows the reduction in skin
  thickness along a perpendicular of the symmetry axis ($K=1$ mode) for
  different deformations of the neutron and proton ground state densities. }
  \label{fig:09}
\end{figure}

Delving deeper into the reason why the $K^{\pi}=1^{-}$ response is reduced with
deformation, it is useful to look at a schematic diagram of the differences
between the spherical and deformed cases.  Fig.~\ref{fig:09} portrays the basic
feature found for the deformed Tin isotopes under study, namely that the
deformation of the proton density is lower than that of the neutron density.
This means that the thickness of the neutron skin is reduced along a
perpendicular of the symmetry axis and this leads to a reduction of the
strength for the $K^{\pi}=1^{-}$ mode. This disturbs the simple 1:2 rule of the
spherical case. This reduction is obviously not fully compensated by a
corresponding increase of the strength of the $K^{\pi}=0^{-}$ mode.  Of course
the details depend on the properties of the orbits occupied by the additional
neutrons, which cannot be explained in this simple picture. 

\begin{figure}[ptbh]
  \includegraphics[width=8.6cm]{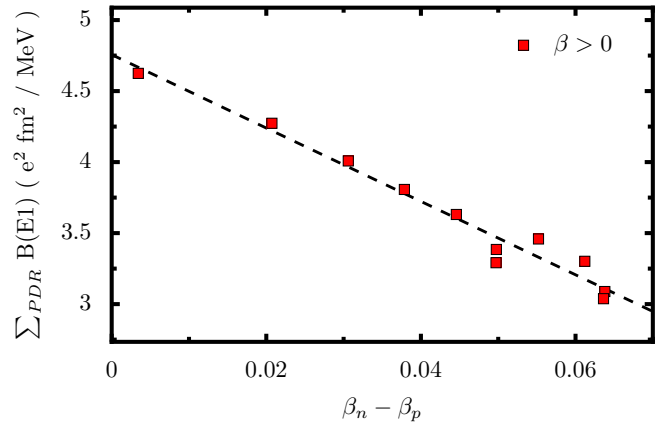}
  \caption{(Color online) Total pygmy strength dependence on the difference of
  deformations for protons and neutrons $\beta_n - \beta_p$ for the deformed
  nuclei in the $^{132-166}$Sn chain. }
  \label{fig:08}
\end{figure}

In Fig.~\ref{fig:08} is plotted the total pygmy strength dependence on the
difference of deformations for the ground state neutron and proton densities
$\beta_{n} - \beta_{p}$. It shows that both are linearly linked. This result is
equivalent to the situation found in spherical nuclei, where the neutron skin
thickness $r_{n} - r_{p}$ determines the total low-lying pygmy strength.
However, in the deformed case, in addition to the difference in neutron-proton
densities radii, the difference of quadrupole deformations $\beta_{n} -
\beta_{p}$ comes into play to determine the total low-lying strength. For the
cases presented in this manuscript, i.e. prolate nuclei, this produces a
reduction in strength in the $K^{\pi}=1^{-}$ mode. For oblate nuclei with
different proton and neutron deformations it is therefore plausible to expect a
similar reduction in overall low-lying strength, caused in this case by a
reduced neutron skin along the symmetry axis and thus a reduced $K=0^{-}$
strength. In summary, the reduced skin thickness along a perpendicular of the
symmetry axis, caused by the different deformations of the neutron and proton
densities, might explain the reduction in strength of the $K^{\pi}=1^{-}$ pygmy
resonance in the deformed Tin isotopes under study.

\section{Conclusions}

The isovector dipole strength in very neutron-rich tin isotopes has been
studied using a fully self-consistent deformed QRPA model based on relativistic
DFT.  The parameter set used in the Lagrangian is NL3~\cite{NL3} with a
monopole force in the pairing channel. The coupling constants of the pairing
interaction were adjusted to reproduce the gap values predicted by the DS1
Gogny force~\cite{Hilaire02}. Due to the weakening of the Z=50 shell closure,
deformation appears for tin nuclei around $^{142}$Sn. In the vicinity of the
N=126 shell closure, beyond $^{164}$Sn, spherical symmetry is restored again.
Therefore axial deformation was included explicitly in these calculations.

The analysis of the evolution of the Giant Dipole Resonance along an isotopic
chain confirms that the Relativistic Quasiparticle Approximation reproduces
basic features predicted by macroscopic hydrodynamical models, namely the
reduction of the centroid position and the splitting of the response in two
modes due to deformation.  However the hydrodynamical models and the RQRPA GDR
position predictions are at variance in the case of very neutron-rich nuclei.
Furthermore, it has been confirmed that th GDR splitting depends linearly on
the deformation.

Regarding the low-lying E1 response, it has been found that deformation hinders
the dipole strength in this region. This effect has been linked to the
suppression of vibrations along a perpendicular of the symmetry axis
($K^{\pi}=1^{-}$ mode) for prolate deformed systems, and explained by the
reduction of neutron skin in this direction caused by the difference in
deformation of the neutron and proton densities.  On the other hand, the
low-lying E1 strength increases with the neutron number, and thus the interplay
of these two effects determines the actual low-lying dipole response in
deformed nuclei.

The analysis of the excitation peaks shows that, even if the low-lying strength
is quenched and spread in deformed nuclei, it nevertheless shows  pygmy
character, with a neutron skin oscillating against neutron-proton core.  The
number of contributing qp-pairs is comparable to that found in the GDR region,
and in agreement with other RQRPA studies in spherical nuclei. It is therefore
concluded that for deformed nuclei with extreme isospin asymmetry the pygmy
mode subsists, but is more spread than in spherical nuclei. Hence, prominent
pygmy modes may be a specific characteristic of spherical neutron rich nuclei
which are not too far from the valley of stability.

\vskip 0.5cm

\noindent{\bf Acknowledgments} The authors wish to thank D. Vretenar
for fruitful discussions. The paper has also been supported by the
Bundesministerium f\"{u}r Bildung und Forschung, Germany under project 06 MT
246 and by the DFG cluster of excellence \textquotedblleft Origin and
Structure of the Universe\textquotedblright\ (www.universe-cluster.de).

\end{document}